\documentclass[12pt]{article}

\usepackage{latexsym}
\usepackage{graphicx}

\begin{document}
\begin{center}
   \textbf{\Large CINEMA AS A TOOL FOR SCIENCE LITERACY}
\end{center}

\begin{center}
\textbf{Costas Efthimiou}\footnote{costas@physics.ucf.edu} and
\textbf{Ralph A. Llewellyn}\footnote{ral@physics.ucf.edu}\\
Department of Physics\\
University of Central Florida\\
Orlando, FL 32816
\end{center}

%%%%%%%%%%%%%%%%%%%%%%%%%%%%%%%%%%%%%%%%%%%%%%%%%%%%%%%%%%%%%%%%%%%%%%%%%
\section{INTRODUCTION}

Surveys conducted by the National Science Foundation (NSF) have thoroughly
documented a severe decline in the understanding of and interest in science
among people of all ages in the United States (NSF, 2002). About 50 percent
of the people do not know that Earth takes one year to complete an orbit
around the sun, that electrons are smaller than atoms, and that early humans
did not live at the same time as the dinosaurs. These examples of faulty
knowledge of physical sciences surely extend to life, social and literary
sciences and are mirrored in other nations.

This paper summarizes an ambitious project embarked upon by the authors at
the University of Central Florida (UCF) to improve public understanding of
the basic principles of physical science, topics often included in the
general education programs of many universities and colleges (Efthimiou {\&}
Llewellyn, 2003). This new approach, \textit{Physics in Films}, uses popular movies
to illustrate the
principles of physical science, analyzing individual scenes against the
background of the fundamental physical laws of mechanics, electricity,
optics, and so on.

While still not a mature project, \textit{Physics in Films} has been successful and has become a
topic of wide discussion, both among UCF students and the physics education
community (APS News, 2003; Chow, 2003; Graham, 2002; Grayson, 2002; Priore,
2003). Only a few similar projects have been tried earlier (Dennis 2002;
Dubeck et al., 1994; Rogers, 2002). None has approached the scope of the
\textit{Physics in Films} project.

%%%%%%%%%%%%%%%%%%%%%%%%%%%%%%%%%%%%%%%%%%%%%%%%%%%%%%%%%%%%%%%%%%%%%%%%%
\section{DESCRIPTION OF THE PROJECT}

\subsection{Genesis of the Physics in Films Project}

It is our experience that students in general think that physics is
difficult, hence boring, and without much relevance in their daily lives.
The authors, who regularly teach physical science for non-science majors,
searched for a way to instill in these students the enthusiasm and
excitement that all physicists experience. After much discussion and review
of existing resources (Dennis, 2002; Dubeck, 1994; Rogers, 2002), the
proposal was made to accomplish this goal by adopting as a teaching vehicle
a medium that students had already accepted as a reflection of today's
culture. The vehicle chosen was the use of popular movies to illustrate both
the basic principles and frontier discoveries of science and also to
motivate students in becoming more critical observers of their world. By
using popular movies as the actual mode of instruction, the intent was to
provide a course in physical science that was more relevant to their daily
lives and to begin to correct the many misconceptions they held about
science.

%%%%%%%%%%%
\subsection{Summer 2002: Action/Adventure Movies}
During the summer 2002 \textit{Physics in Films} was offered for
the first time. Principles of physics were discussed using scene
clips from nine popular action/adventure movies. For example, the
law of gravitation as (mis)used in ``Independence Day,''
conservation of momentum in ``Tango {\&} Cash,'' speed and
acceleration in ``Speed 2'' and so on.

Figure 1 shows the nine movies used that first summer. Students were
required to watch the films at home and turn in a brief, written analysis of
the physics principle illustrated in each of three scenes of their own
choosing (homework!). In class, five to ten percent of the class (of 90
students) were called upon each day to orally present their analysis of one
scene to the class. Both the written and oral analyses became part of their
grade in the course.

\begin{figure}[h!]
\begin{center}
\includegraphics[width=3.5in]{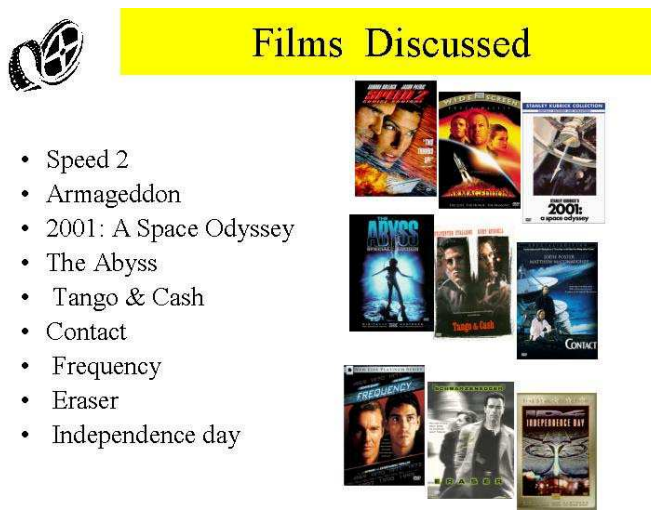}
\end{center}
\caption{Summer 2002: Action/Adventure/SciFi Movies.}
\label{fig:1}
\end{figure}

\begin{figure}[h!]
\begin{center}
\includegraphics[width=3.5in]{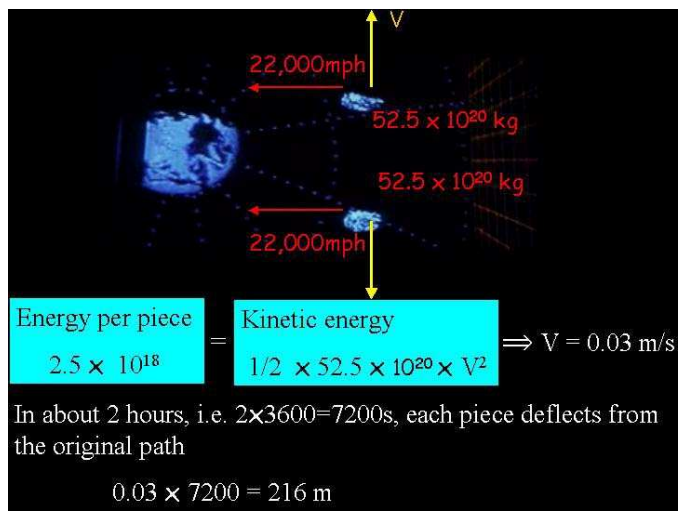}
\end{center}
\caption{Asteroid Pieces Approach Earth}
\label{fig2}
\end{figure}

\textbf{\textit{An Example: Armageddon.}} As an example of how a
movie clip is used in illustrating a physical principle, consider
``Armageddon'' (starring Bruce Willis). A huge, errant asteroid
the size of Texas is on a collision course with Earth. (There are
no known asteroids that large.) A team of oil well drillers is
dispatched via a pair of space shuttles to intercept the asteroid,
drill a hole in it at the right place, lower a large nuclear bomb
into the hole, and subsequently blow the asteroid into two large
pieces. The transverse velocities imparted to the two pieces by
the explosion, when added to their (undiminished) velocities
toward Earth, are to cause the pieces to just miss Earth, thereby
averting worldwide disaster. After showing the clip, the analysis
uses conservation of energy, conservation of momentum, vector
addition, and the law of gravity to assess how physically
realistic the solution in the film (and in such a future event)
might be. The overall situation is depicted in Figure \ref{fig2}.

Using numbers provided in the film, the students are introduced to the idea
of making reasonable approximations. For example, the asteroid is the size
of Texas, so Texas is assumed to be a square whose surface area equals that
of the state. Then the asteroid is approximated as a cube, each of whose
sides equals the surface area of the state. Multiplying the volume of the
cube by the average density of Earth gives us a decent estimate of the mass
of the asteroid.

Assuming the nuclear bomb to be equal to 100,000 Hiroshima bombs provided an
estimate of the energy available for the job (a modern nuclear warhead
equals about 1,000 Hiroshima bombs). Assuming all of that energy became
kinetic energy equally divided between the two pieces of the asteroid (i.e.,
ignoring the energy needed to break the asteroid into two pieces), we can
readily compute the distance the pieces have moved perpendicular to their
original direction of motion by the time they reach Earth. As noted in the
diagram, that distance is only a bit over 200 meters for each piece.

The students are astonished. Instead of being hit by one Texas-size
asteroid, Earth will be hit by two half-Texas-size asteroids about 400
meters apart! This discussion concludes with an explanation of what is
realistically possible and why the government has an ongoing project to
detect and track space objects approaching Earth or in Earth-crossing orbits
around the sun.

\subsection{Personal Response System}
 From the beginning all
sections of \textit{Physics in Films} have used a personal
response system that enables each student to register their
attendance and to record answers to questions asked in class by
the instructor. The system provides immediate confirmation of
answers, permits the students to change their answers, displays
the correct answers, and provides a histogram of the class
responses to questions so that students can compare themselves to
the class as a whole.

Attendance records and responses to quiz questions, both of which contribute
to the final grade, are recorded automatically by an in-class computer. The
system also provides a means of recording student opinions regarding various
aspects of the course, information that is helpful in improving the
presentations. For example, Figure 3 contains two tables showing student
responses to the use of the system itself. (SA=strongly agree, A=agree, N=no
opinion, D=disagree, SD=strongly disagree.) Shown, too, in the picture are a
few communication pads like the ones used by individual students.

\begin{figure}[h!]
\begin{center}
\includegraphics[width=3.5in]{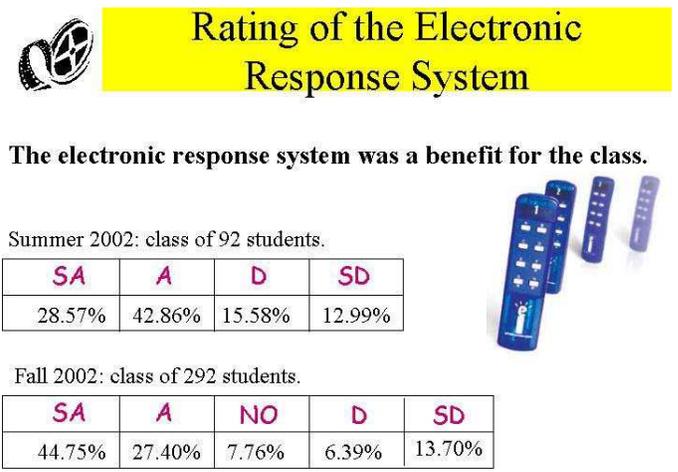}
\end{center}
\caption{Rating the Electronic Response System}
\label{fig3}
\end{figure}

%%%%%%%%%%%
\subsection{Development of Flavors}

After \textit{Physics in Films }had been given for three terms to four sections with a combined
enrollment of about 800, the improved performance of the students relative
to the traditional physical science course together with their enthusiasm
regarding the \textit{Physics in Films} approach motivated the authors to explore new directions for
developing the course further. The original syllabus included movies
selected to span the entire topical range of the traditional physical
science course. No special attention was given to the genre or theme of the
films used. The films eventually used, like those in Figure 1 for the first
term, were action/adventure and science fiction films. Encouraged by the
students' enthusiasm, the authors considered possible variations of the
course that would accommodate the curiosity of every student and satisfy the
needs of every instructor. It was decided to create versions
(packages)---nicknamed ``flavors''---whereby each flavor used films of a
particular genre or theme. Plans have been developed to create the following
flavors: Action/Adventure, Science Fiction, Superhero, Modern Physics,
Astronomy, Pseudoscience, and Metaphysics. During the summer and fall 2003
terms the Superhero and Pseudoscience flavors were given for the first time.

\subsection{Physics in Films: Superheroes}
Part of the motivation for offering the Superheroes flavor came
from a course given by Jim Kakalios at the University of
Minnesota-Twin Cities (Feder, 2002). He has taught a successful
course in physical science based on superhero comic books. The
\textit{Physics in Films: Superheroes} flavor complements
Kakalios' approach, substituting motion and `real' action for
static images. It was first taught in the summer 2003 term using
the films shown in Figure 4.

\begin{figure}[htb!]
\begin{center}
\includegraphics[width=3.5in]{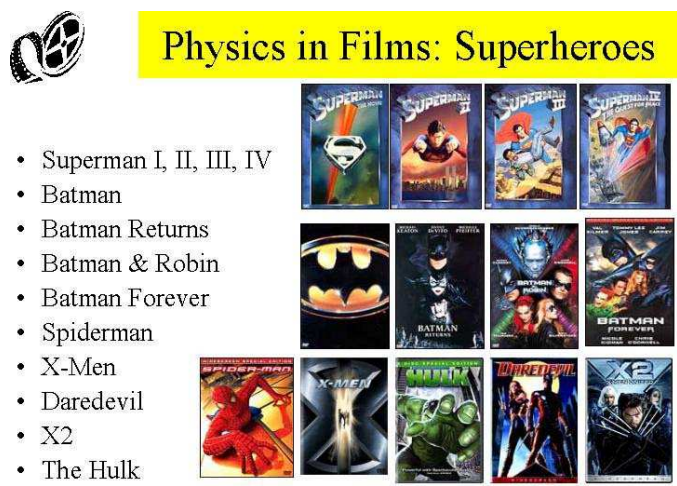}
\end{center}
\caption{Physics In Films: Superheroes Films}
\label{fig4}
\end{figure}

When the course was in progress, two of the films (``The Hulk''
and ``X2'') were not yet available on disk or tape, so the
students had to go to the theaters, but no one complained. In
fact, the whole class approved!

The topical content of \textit{Physics in Films} originally closely followed that of traditional
physical science courses. The textbook was ``Fantastic Voyages'' (Dubeck et
al., 1994). While somewhat `lighter' than a typical physical science
textbook, it uses some (rather old) movies as illustrations and is the only
such text available. In \textit{Physics in Films:} \textit{Superheroes} classes however, the authors deviated somewhat
from the traditional path and added some decidedly non-traditional books.
They were ``The Science of Superheroes'' (Gresh {\&} Weinberg, 2002), ``The
Science of Superman'' (Woolverton, 2003), and ``The Science of XMen'' (Yaco
{\&} Haber, 2000), all illustrated in Figure 5. Stocked in the trade book
sections of the bookstores, they are replete with applications of the
concepts of physical science. The authors of these books frequently use
physical laws in an effort to make the powers of the superheroes plausible
to the lay reader. In so doing they provide fertile ground for lively
discussions.

\begin{figure}[htb!]
\begin{center}
\includegraphics[width=3.5in]{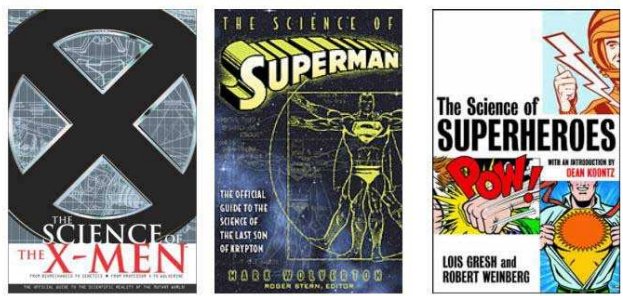}
\end{center}
\caption{\textit{Physics In Films: Superheroes} books.}
\label{fig5}
\end{figure}

\textbf{\textit{An Example: ``Spiderman}}.'' As an illustration of a
superhero film used in the course consider ``Spiderman'' (starring Tobey
McGuire) and a specific scene. The scene is where the Green Goblin is
standing on one of the towers of the Queensboro Bridge in New York
simultaneously holding Mary Jane (MJ) suspended over the river with his left
hand and the broken cable holding the Roosevelt Island cable car filled with
children with his right hand. This one scene provides for discussions in
equilibrium of forces, torque, friction, and free fall. Considering the
latter, the Green Goblin presents Spiderman, who is crouched on the bridge
superstructure, with a dilemma by allowing both MJ and the cable car to
fall. Which will Spiderman save?

Timing events directly from the film clip, students see that it takes
Spiderman 14 seconds to decide and start after MJ first, saving her, then
saving the children in the cable car. It looks great in the film! But when
the instructor leads the students through analysis of MJ's fall they
discover that in those 14 seconds she would fall a distance $d$ =
$\raise.5ex\hbox{$\scriptstyle 1$}\kern-.1em/
\kern-.15em\lower.25ex\hbox{$\scriptstyle 2$}  \quad g \quad t^{2}$ where $g$ = 9.8
m/s$^{2}$ and $t$ = 14 s. Thus, d = 960 meters and the Queensboro Bridge is
only 106 meters above the water! Once again, the students are astonished!
Even assuming that the director implies `slow motion' effects as he presents
the events, we can estimate that Spiderman cannot react and catch MJ in less
than 5 seconds. This would give a free-fall length of 122.5 meters, still
more than the height of the bridge.

\subsection{Physics in Films: Pseudoscience}
 An idea or theory is
called pseudoscience if it contradicts accepted scientific data,
but is presented as scientific. Note that a mistake or error in
presenting scientific data does not signal pseudoscience. It is
the intentional misrepresentation of facts or unverified claims
that justify the label. For our purposes the authors categorize as
pseudoscientific those movies that are based on topics or
phenomena that contradict scientific facts. There are many such
films that might be used, but a group was chosen that most
students had already seen or knew about. (See Figure 6.)
\textit{Physics in Films: Pseudoscience} was first taught in the
summer 2003 term.

\begin{figure}[h!]
\includegraphics[width=4in]{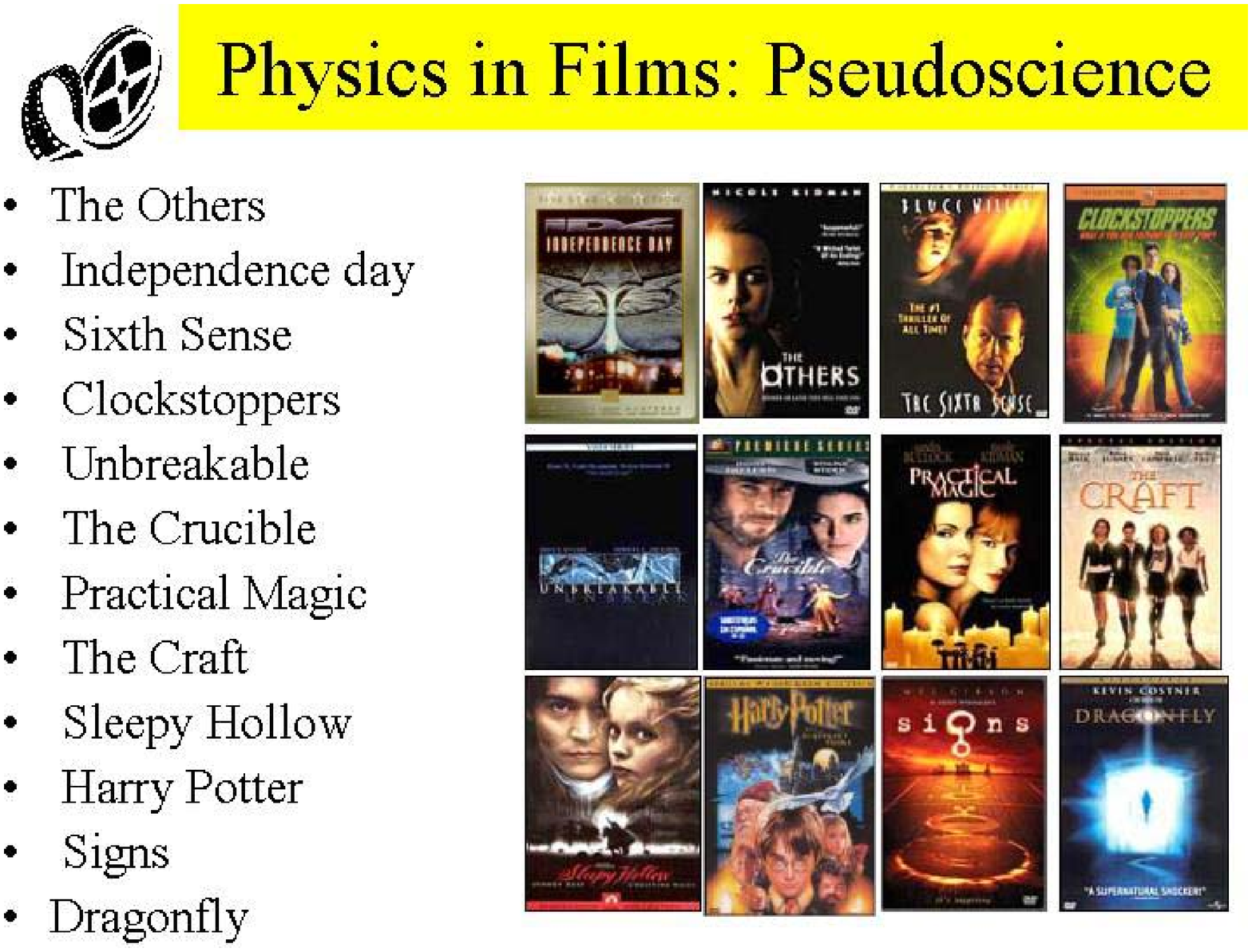}
\caption{Films used in Physics in Films: Pseudoscience}
\label{fig6}
\end{figure}

The topics covered and related to pseudoscience included, among others:

\begin{itemize}
\item Universality of physical laws -- magic
\item Time -- time reversal, time stopping
\item Strength of materials -- unbreakability
\item Chemical reactions -- zombies
\item Fundamental interactions -- ghosts
\item Intelligent life in the universe -- alien visitors to Earth, alien abductions
\end{itemize}

\begin{figure}[h!]
\begin{center}
\includegraphics[width=4in]{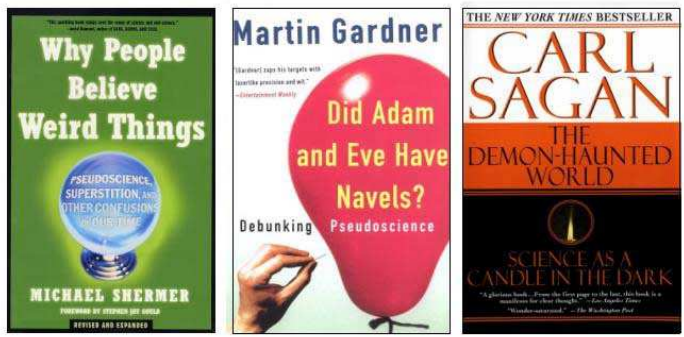}
\end{center}
\caption{Bookss used in \textit{Physics in Films: Pseudoscience}.}
\label{fig7}
\end{figure}

\textbf{\textit{An Example: The Sixth Sense}}. Among the topics discussed in
physical science are the related concepts of temperature and heat transfer
via conduction, convection, and radiation. ``The Sixth Sense'' (starring
Bruce Willis) is concerned with ghosts. The movie tells us that ghosts like
low temperatures, although why that should be is not made clear. In the
scene where the young hero goes to the bathroom during the night, a sudden
drop in temperature is clearly shown. We have been told that this heralds
the appearance of a ghost, and indeed one appears.

As it happens, a scientific study of this presumed phenomenon has been done
(Frood, 2003). The Haunted Gallery of Hampton Court Palace near London, UK,
is reported by many visitors to be haunted by the ghost of Catherine Howard
(fifth wife of Henry VIII, executed in 1542). Air motion and thermal
detectors were deployed in the Gallery and some 400 visitors were asked
about their experiences during the visit. More than half felt sudden drops
in temperature, some sensed ghostly presence, and several reported seeing
Elizabethan figures. The study (Frood, 2003) revealed many poorly sealed
hidden doorways that admitted columns of colder exterior air. In two
locations the temperature of the localized draft was only 36$^{o}$ F!

\textbf{\textit{A Course Book Display.}} The UCF campus bookstore cooperated
by installing a prominent display `island' featuring the \textit{Physics in Film: Superheroes} and \textit{Physics in Film: Pseudoscience} courses
(Figure 8). The display attracted student attention and (the authors
suspect) increased sales of the displayed books, as well.

\begin{figure}[h!]
\begin{center}
\includegraphics[width=1.50in]{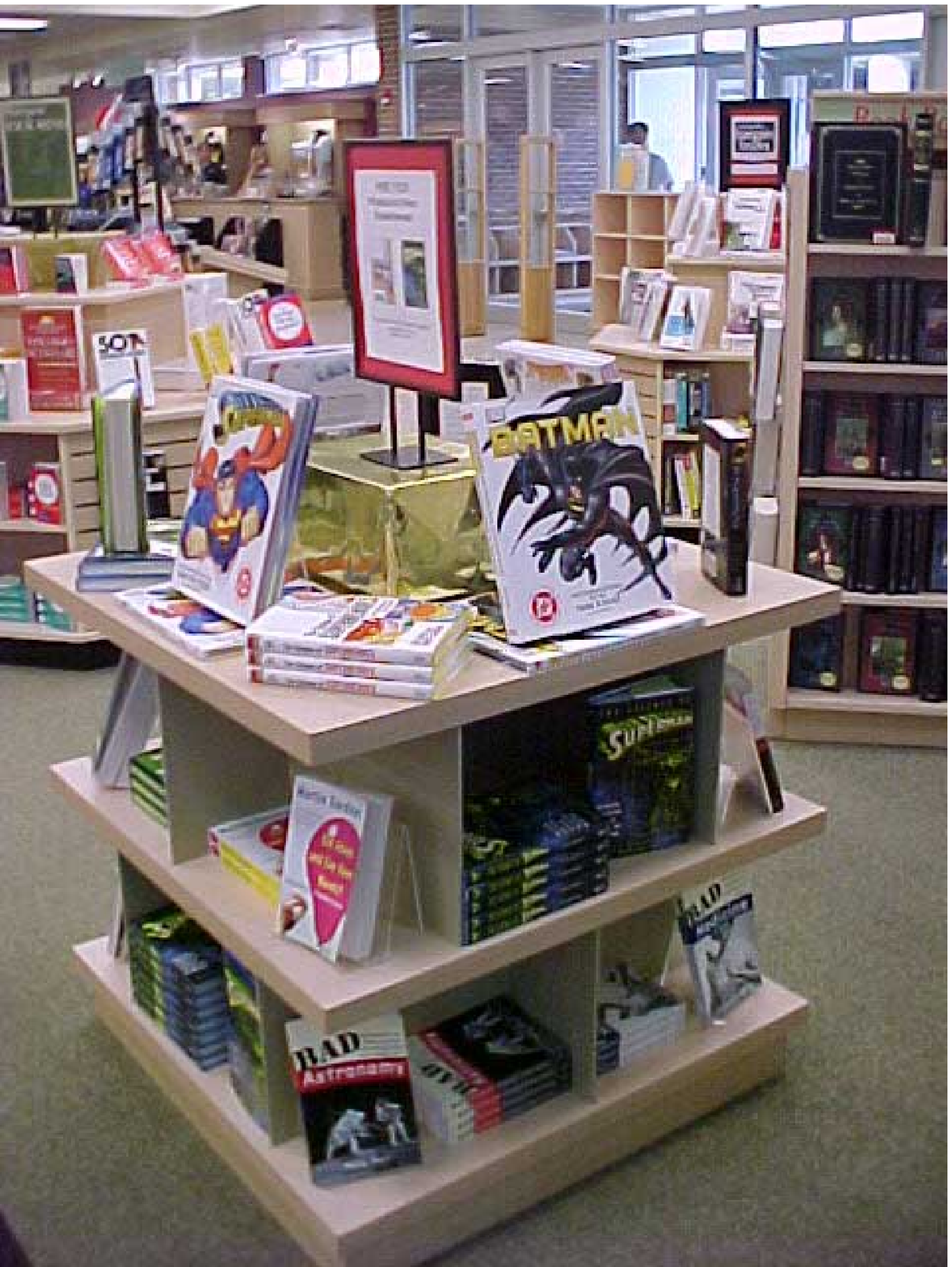}
\includegraphics[width=1.50in]{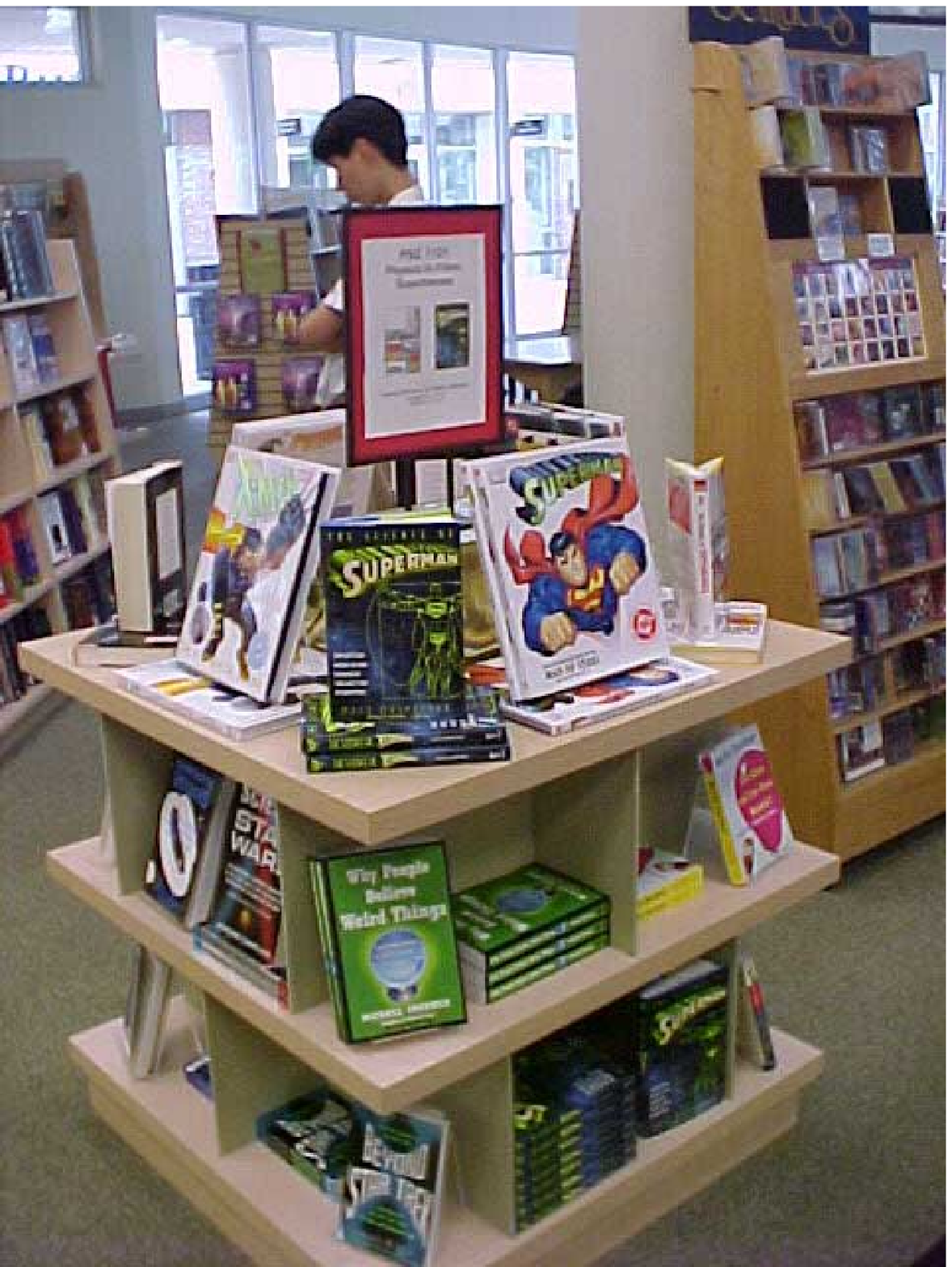}
\end{center}
\caption{UCF Bookstore display for \textit{Physics in Films}.}
\label{fig8}
\end{figure}

%%%%%%%%%%%%%%%%%%%%%%%%%%%%%%%%%%%%%%%%%%%%%%%%%%%%%%%%%%%%%%%%%%%%%%%%%%%%%%%%%%%%%%%%%%%
\section{ASSESSMENT AND FINDINGS}

As of this writing, the \textit{Physics in Films} alternative to the traditional Physical Science
course has been offered at UCF since the summer 2002 term to a total of nine
sections enrolling, collectively, about 1600 students. During that time the
authors have collected data regarding student evaluations of the course and
their performance on exams. In that same time period there have been seven
sections of Physical Science taught in the standard way, enrolling about
2000 students. Some of the more interesting results and comparisons are
presented in this section.

%%%%%%%%%%%%%%%%%%%%%%%%%%%%%%%%
\subsection{Examination Results}

Even though students may embrace a new idea enthusiastically, that does not
mean their performance will necessarily improve relative to the traditional
course (where most of them really struggle). In fact, student performance on
individual exams and overall is improved in the \textit{Physics in Films} sections when compared with
those in traditional Physical Science sections. Table 1 shows the exam
scores distribution for two classes of about the same size (295 students
each), covering the same topics, and taught by the same instructor. The
results are obviously dramatically different!

\begin{table}[h!]
\begin{center}
\begin{tabular}{|c|c|c|c|c|} \hline
Instructional Mode& Exam 1& Exam 2& Exam 3& Final Exam
     \\ \hline
Traditional  &  & & & \\
Class average percent & 49.3& 65.3& 58.2& 59.4 \\ \hline
\textit{Physics in Films} & & & & \\
Class average percent & 4.9& 67.7& 75.7& 72.8 \\ \hline
\end{tabular}
\end{center}
\caption{Comparison of Exam Results for large enrollment sections.
         The sections
         were taught by the same instructor on the same material.}
\label{tab1}
\end{table}

%%%%%%%%%%%%%%%%%%%%%%%%%%%%%
\subsection{Student Evaluations and Opinions}

Since Physics in Films was first offered, considerable data has been
collected regarding student opinions and evaluations of the course and the
various flavors taught to date. This section summarizes some of the more
interesting of those data.

\begin{table}[h!]
\begin{center}
\begin{tabular}{|c|c|c|c|c|c|}
\hline Term & Strongly  Agree& Agree & No Opinion& Disagree&
      Strongly  Disagree \\ \hline
Summer 2002 & & & & & \\
 92 students & 77.9{\%}& 10.4{\%}& n/a& 9.1{\%}&2.6{\%} \\ \hline
Fall 2002  & & & & & \\
 292 students & 56.9{\%}& 26.6{\%}& 6.9{\%}& 4.1{\%}& 5.5{\%}   \\ \hline
\end{tabular}
\end{center}
\caption{Responses to the Statement: \textit{Physics in
          films should be developed further since it is more interesting
          than the standard physical science course.}}
\label{tab2}
\end{table}

\begin{table}[h!]
\begin{center}
\begin{tabular}{|c|c|c|c|c|} \hline
Strongly  Agree& Agree & No Opinion& Disagree& Strongly  Disagree
   \\ \hline
60{\%}& 32{\%}& 5{\%}& 1{\%}& 2{\%} \\ \hline
\end{tabular}
\end{center}
\caption{Response to the statement: \textit{I think I learned
         something from this class (Physics in Films: Superheroes)}.}
\label{tab3}
\end{table}

\begin{table}[h!]
\begin{center}
\begin{tabular}{|c|c|c|c|c|}\hline
Strongly   Agree& Agree & No  Opinion& Disagree& Strongly Disagree
\\ \hline
68{\%}& 25{\%}& 5{\%}& 2{\%}& 0{\%} \\ \hline
\end{tabular}
\end{center}
\caption{Responses to the statement: \textit{I would recommend to
         my friends that they take this course (Physics in Films:
         Pseudoscience)}.}
 \label{tab4}
\end{table}

\begin{table}[h!]
\begin{center}
\begin{tabular}{|c|c|c|c|c|} \hline
Strongly  Agree& Agree & No  Opinion& Disagree& Strongly  Disagree \\
  \hline
42{\%}& 42{\%}& 16{\%}& 0{\%}& 0{\%} \\ \hline
\end{tabular}
\end{center}
\caption{Responses to the statement: \textit{The topics selected
         from the movies for physics analysis were interesting}.}
\label{tab5}
\end{table}

However, changing the public's overall perception of science is not easy.
More and longer term efforts reaching a much broader audience will be
needed. As Table 6 suggests, fear and unreasonable dislike of science are
deeply rooted in the minds of students and others, as the NSF surveys cited
in the introduction to this paper makes clear.

\begin{table}[h!]
\begin{center}
\begin{tabular}{|c|c|c|c|c|} \hline
Strongly  Agree& Agree & No  Opinion& Disagree& Strongly  Disagree \\
   \hline
25{\%}& 26{\%}& 8{\%}& 18{\%}& 23{\%} \\ \hline
\end{tabular}
\end{center}
\caption{Responses to the statement: \textit{I do not like science
         and I do not want to read anything about science once I have
         finished this course}.}
\label{tab6}
\end{table}

%%%%%%%%%%%%%%%%%%%%%%%%%%%%%%%%%%%%%%%%%%%%%%%%%%%%%%%%%%%%%%%%%%%%%%%%%%%%%
\section{CONCLUSIONS AND RECOMMENDATIONS}

It appears clear from the information and data presented herein that the
\textit{Physics in Films} alternative to the more traditional Physical Science course captures
student interest and improves their performance. Approximately half of the
students who enroll in Physical Science at UCF now take the \textit{Physics in Films} version, a
further testament to the success of the approach.

%%%%%%%%%%%%%%%
\subsection{The Future}

The authors will certainly continue their work at UCF in further developing
the \textit{Physics in Films} concept. However, it may be argued that will not be enough. As Table 6
and the NSF surveys have made clear, changing public perceptions of science
will be neither easy nor quickly accomplished.

It is the goal of the authors to increase the awareness of science
and to show that an understanding of basic physical science can be
both enriching and rewarding. To this end they are working toward
the development of `packaged' \textit{Physics in Films} flavors
that can be readily transferred to other institutions. They are
also writing a new physical science textbook designed to support
the \textit{Physics in Films} mode of teaching. In addition, they
have begun to explore the application of the concept to the
creation or enhancement of general education courses in many other
disciplines. The following list  presents several possibilities
together with a few examples of films that include material for
each discipline.

\begin{itemize}
\item \textbf{Mathematics in Films:} \textit{Pi, Good Will Hunting, Pay it Forward, Contact}
\item \textbf{Astronomy/Astrophysics in Films: } \textit{Armageddon, Deep Impact, Contact}
\item \textbf{Biology in Films:} \textit{Spiderman, The Hulk, Planet of the Apes, Jurassic Park}
\item \textbf{Chemistry in Films:} \textit{Flubber, Year of the Comet}
\item \textbf{Engineering in Films}: \textit{Armageddon, The Bridge on the River Kwai}
\item \textbf{Archeology/Anthropology in Films:} \textit{Indiana Jones trilogy, Jurassic Park}
\item \textbf{Computers in Films}: \textit{The Net, Independence Day, War Games}
\item \textbf{Philosophy in Films:} \textit{Blade Runner, Matrix, Terminator trilogy, Ghost}
\item \textbf{History in Films}: \textit{Braveheart, Patriot, The Man in the Iron Mask}
\item \textbf{Law in Films}: \textit{Erin Brockovich, The Firm, Legally Blond, Primal Fear}
\item \textbf{Forensic Science in Films:} \textit{Jennifer 8, Murder by the Numbers, Bone Collector}
\end{itemize}

%%%%%%%%%%%%%%%%%%%%%%%%%%%%%%%%%%%%%%%%%%%%%%%%%%%%%%%%%%%%%%%%%%%%%%%%%%%%%%%%%%%%%%%%%%%%%%%%

\end{document}